# VaxPulse: Active Global Vaccine Infodemic Risk Assessment

Gerardo Luis DIMAGUILA[a,b,c,1], Muhammad JAVED[a,b,c,d], Jeremiah MUNAKABAYO[a,b], Sedigh KHADEMI[a,b,c], Hazel CLOTHIER[a,b,c,d], Joanne HICKMAN[a], Jim BUTTERY[a,b,c,d,e]

[a] *Epidemiology Informatics, Centre for Health Analytics, Melbourne Children's Campus, Australia*
[b] *Surveillance of Adverse Events Following Vaccination In the Community (SAEFVIC), Murdoch Children's Research Institute, Australia*
[c] *Department of Paediatrics, The University of Melbourne, Australia*
[d] *Global Vaccine Data Network, University of Auckland, New Zealand*
[e] *Infectious Diseases, Department of General Medicine, Royal Children's Hospital, Australia*

ORCiD ID: Gerardo Luis Dimaguila https://orcid.org/0000-0002-3498-6256

**Abstract.** Vaccine infodemics, driven by misinformation, disinformation, and inauthentic online behaviours, pose significant threats to global public health. This paper presents our response to this challenge, demonstrating how we developed VaxPulse Vaccine Infodemic Risk Assessment Lifecycle (VIRAL), an AI-powered social listening platform designed to monitor and assess vaccine-related infodemic risks. Leveraging interdisciplinary expertise and international collaborations, VaxPulse VIRAL integrates machine learning methods, including deep learning, active learning, and data augmentation, to provide real-time insights into public sentiments, misinformation trends, and social bot activity. Iterative feedback from domain experts and stakeholders has guided the development of dynamic dashboards that offer tailored, actionable insights to support immunisation programs and address information disorder. Ongoing improvements to VaxPulse will continue through collaboration with our international network and community leaders.

**Keywords.** Public Health Informatics, Vaccination Hesitancy, Learning Health System, Machine Learning, Information Systems, Public Health Surveillance

## 1. Introduction

Vaccine infodemics, characterised by an overwhelming amount of information – including gaps, concerns, mistrust, and exposure to misinformation and disinformation – pose significant challenges to global public health [1]. Increasingly risks of adverse events following immunisation (AEFI) are viewed as more serious than disease risks, exacerbated during the pandemic [2]. The rapid spread of vaccine-related misinformation through online platforms has undermined vaccine confidence, leading to decreased immunisation rates and outbreaks of vaccine-preventable diseases, at enormous cost to lives and economies [3, 4]. The issue of misinformation and disinformation has prompted a legislative response in Australia, with a bill introduced on 12 September 2024 to combat what it calls 'information disorder' [5]. The bill focused on mitigating serious harm, including harm to public health, and to the Australian economy. It also highlights the concept of "inauthentic behaviour," coordinated actions that amplify misinformation regardless of the content's truthfulness. In Indonesia, despite a strong history of vaccine development, the pandemic caused a drop in confidence and thus vaccine coverage and confidence, with the proportion of unvaccinated infants rising from 10% to 26% in 2021 due to fears about multiple immunisations and adverse events following immunisation (AEFI) [6]. The COVID-19 pandemic was not the first to have an associated vaccine-related infodemic; in 2017, misinformation around the dengue vaccine in the Philippines led to a sharp decline in vaccine confidence [7].

---

[1] Corresponding Author: Gerardo Luis Dimaguila, gerardoluis.dimaguil@mcri.edu.au

Our aim is to address the question of how government agencies and vaccine advocacy groups can effectively detect and respond to large-scale, online infodemic risks that jeopardise the success and resilience of immunisation programs. We developed VaxPulse, a platform for infodemic detection and response. The global scope of vaccine-related infodemics underscores the urgent need to identify, develop, and rigorously evaluate strategies for effective management.

## 2. Methods

Developing VaxPulse was an iterative process of discovery and refinement. Social listening, a common marketing strategy, analyses social media data to identify online trends. In public health, it traditionally relies on offline sources like health systems, communities, and expert studies [8]. During COVID-19, the World Health Organization (WHO) used online social listening to track global narratives and assist infodemic managers. Their platform, called EARS, categorised social media data (December 2020- February 2022) by themes, like the virus' origins, symptoms, and treatments [8].

We also established VaxPulse during the pandemic, and have continually refined it over several years, incorporating lessons learnt along the way. From the outset, it was conceptualised as an AI social listening learning health system (LHS) [9] monitoring sentiments and concerns [10] about vaccine safety and personally experienced vaccine reactions [11] (ethics: HREC/85026/RCHM-2022). Insights were used with heuristic evaluation to adapt a jurisdictional public-facing COVID-19 vaccine safety report [9]. Importantly, the LHS established a collaboration between end users and domain experts in consumer and public engagement, communication and social science, machine learning, immunology and infectious diseases, and vaccine safety and epidemiology [9].

Analysing social media data led us to encounter the presence of online inauthentic behaviour in the form of social bots. Social bots are partially or fully automated accounts that post and engage with, including resharing, social media content [12]. In 2021, we used a novel machine learning technique to analyse COVID-19 online vaccine safety discourse, revealing that social bots heavily influenced human sentiment and hesitancy, often amplifying vaccine-related concerns [12]. Our award-winning method, recognised at Australia's Health Informatics Knowledge Management Conference, is being refined to counter advanced GenAI-powered social bots. Funding received at the 2024 National Communicable Diseases and Immunisation Conference supports this enhancement.

As we continued to develop VaxPulse, it became apparent that monitoring misinformation and disinformation were important in managing the information disorder, and future infodemics. The pandemic significantly increased the proliferation of misinformation and disinformation [8], and this was evident in international meetings, conferences, and seminars we contributed and participated in. We soon developed a process to monitor information disorder. Our use of deep learning approaches with active learning and data augmentation allows for more accurate and real-time analysis of online sentiments and misinformation trends. Evaluation of our models are reported elsewhere.

We have since expanded our local collaboration for continued development to include additional expertise, forming an international network of experts and national vaccine administration officials from Indonesia, Philippines, India, Canada, Switzerland, and the World Health Organization. By partnering with WHO, we will have access to the WHO EARS data, for integration with VaxPulse. Through this process of identifying and developing critical components to assess infodemic risks, illustrated in Figure 1, we have established VaxPulse Vaccine Infodemic Risk Assessment Lifecycle (VIRAL), an ongoing, active surveillance and assessment of global infodemic risks that may threaten immunisation programs.

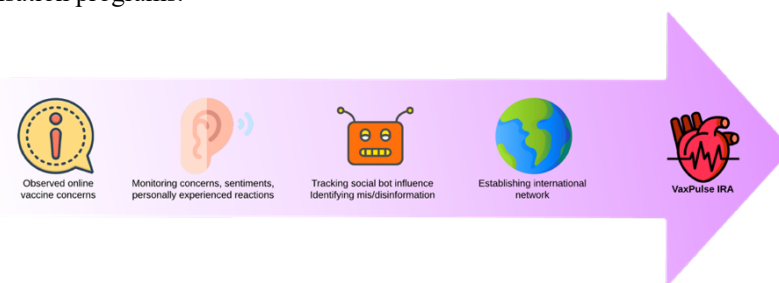

**Figure 1.** Process of identifying and developing VaxPulse VIRAL components.

We presented the components of VaxPulse VIRAL to the Asia-Pacific Vaccine Research Network (APVRN) [13], comprised of National Immunisation Technical Advisory Group members, vaccinologists, vaccine researchers, clinicians and Ministry of Health personnel from across seven countries in the region, to gather feedback and insight into what VIRAL components would be most informative for their work. To present VIRAL insights across various vaccines, we developed data visualisation dashboards through PowerBI, a business intelligence tool. The choice of what VIRAL components to prioritise in the landing page, and which visualisations would be necessary were determined through the process above, and through feedback of APVRN. GLD, an informatics specialist developed draft wireframes, and was reviewed by JB an immunologist and paediatrician, MJ a machine learning expert, and JM a data scientist. Initial iterations of the dashboard were then presented to a group for review and feedback comprising of experts in epidemiology, vaccine safety, paediatrics, immunology, infectious disease, consumer engagement, data science, and machine learning. We intend to continue gathering feedback from stakeholders and domain experts.

## 3. Results

VaxPulse VIRAL continuously monitors and provides insights on public sentiments of priority vaccines and immunisation programs; topics of vaccine concerns and self-reported adverse events; influence of social bots in online conversations and spread of sentiments; vaccine mis-/disinformation in online discussions, posts, and media; and complexity of language vis-à-vis sentiments and mis-/disinformation. We aim to calculate Infodemic Risk Index to quantify exposure to unreliable news in future [14].

VaxPulse VIRAL provides timely data insights on all known infodemic forms by applying innovative machine learning methods that incorporate fine-tuned classifiers and GenAI to collect, augment and process data. Online media will be prioritised using in-country local knowledge. These dashboards could be used as a "deck of graphs", a set of VIRAL insights updated regularly. They would produce better understanding of how information flows on social networks, and the relationship between online engagement, belief change, and perceptions of vaccines. Timely VIRAL data insights would be provided to government agencies, infodemic managers, and vaccine uptake groups to allow them to respond to public vaccine concerns – which could put immunisation programs at risk – faster and more effectively. This would also ensure healthcare workers are informed of real-time concerns and equipped with accurate information for effective conversations. Figure 2 presents the landing page of VaxPulse VIRAL.

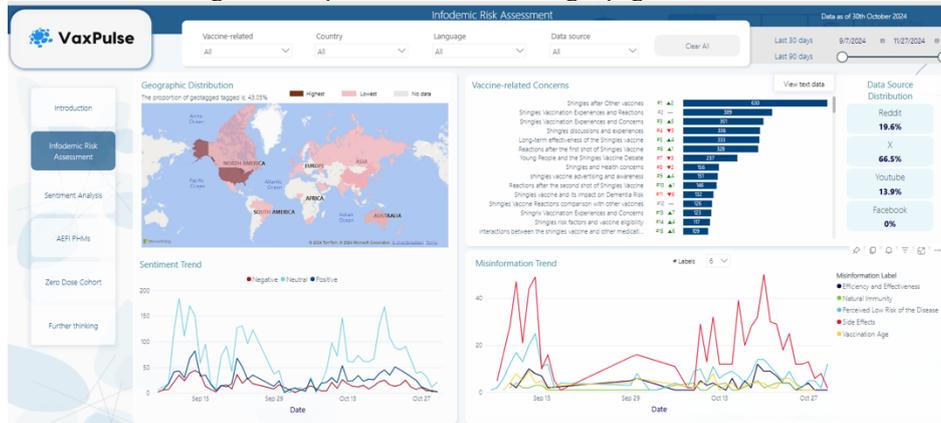

**Figure 2.** VaxPulse Infodemic Risk Assessment Dashboard.

## 4. Discussion

Vaccine infodemics are an ongoing critical challenge, making their effective management an urgent priority. To address this challenge, we developed the VaxPulse Vaccine Infodemic Risk Assessment Lifecycle, that will act as a nucleus for an information disorder mission control virtual environment, providing timely insights and enabling relevant groups to take action in improving vaccine uptake.

To generate a near real-time 'pulse' of infodemic risks, we are collaborating with country partners to produce regular, actionable insights tailored to their vaccine programs. VIRAL dashboard could be stratified along geographic communities. We are developing multilingual support, starting with Filipino, Urdu, Hindi, Bangla, Spanish, and Farsi, to allow for ethno-lingual stratification of VIRAL. This would allow infodemic, social science, and vaccine uptake groups to implement localised strategies to tackle infodemics, misinformation, and vaccine hesitancy. We will work with local leaders to refine LLM strategies, online sources, and the relevance of dashboards for actionable insights. We also intend to develop tested response strategies based on VaxPulse VIRAL in a LHS.

Currently, the dashboard highlights the top concerns for the period, indicating which concerns may require a response to mitigate the risk of declining public vaccine confidence and to inform vaccine safety communication strategies. To identify emerging trends, we initially experimented with tracking top concern changes over consecutive periods, but quantifying the significance of these changes proved challenging. We are now trialling adapting a formula for calculating 'velocity change' of topic concerns [8].

## 5. Conclusions

VaxPulse VIRAL and its components were identified and developed over several years through lessons learnt, interdisciplinary expertise, international engagement, and a review process. Interrogatable dashboards will allow for prioritisation of local themes of clinical and social importance, providing direct insights to immunisation programs and local opinion leaders. This will enable timely analysis of vaccine-specific infodemic risks for public and healthcare workers and create an accessible dashboard resource for local leaders. We will continue to improve VaxPulse by working with our international network and community leaders, and we remain open to future collaborations.